\documentclass[fleqn,usenatbib]{mnras}

\usepackage{newtxtext,newtxmath}
\usepackage[T1]{fontenc}

\DeclareRobustCommand{\VAN}[3]{#2}
\let\VANthebibliography\thebibliography
\def\thebibliography{\DeclareRobustCommand{\VAN}[3]{##3}\VANthebibliography}

\usepackage{graphicx}	% Including figure files
\usepackage{amsmath}	% Advanced maths commands

\title[Pa 30 Morphology via Cooling]{Sculpting the Morphology of Supernova Remnant Pa 30 via Efficient Ejecta Cooling}

\author[Duffell, Polin, and Mandal]{
Paul C. Duffell,$^{1}$\thanks{E-mail: pduffell@purdue.edu}
Abigail Polin$^{1,2,3}$
and Soham Mandal$^{1}$
\\
$^{1}$Department of Physics and Astronomy, Purdue University, 525 Northwestern Avenue, West Lafayette, IN 47907, USA\\
$^{2}$The Observatories of the Carnegie Institution for Science, 813 Santa Barbara St., Pasadena, CA 91101, USA\\
$^3$TAPIR, Walter Burke Institute for Theoretical Physics, 350-17, Caltech, Pasadena, CA 91125, USA\\
}

\date{Accepted XXX. Received YYY; in original form ZZZ}

\pubyear{2015}

\begin{document}
\label{firstpage}
\pagerange{\pageref{firstpage}--\pageref{lastpage}}
\maketitle

\begin{abstract}
We demonstrate in a proof-of-concept numerical hydrodynamics calculation that the narrow radial filamentary structures seen in Pa 30 could be generated through highly efficient cooling (e.g. via line emission) in the ejecta.  Efficient cooling in the ejecta causes a drop of pressure support in Rayleigh-Taylor fingers, leading them to be compressed, and suppressing the growth of Kelvin-Helmholtz instability.  Following this result, we make three predictions that could determine whether this is the mechanism responsible for shaping Pa 30:  First, we predict very strong emission lines, strong enough to cool a significant fraction of the shock energy in an expansion time.  Secondly, we predict that the forward shock should be highly corrugated on small scales, with the shock front closely following the structure of the filaments.  Third, we predict that these filaments should be nearly ballistic, with velocities around 90\% of the free-expansion velocity ($v \approx 0.9 ~r/t$).  These predictions should be falsifiable in follow-up observations of this remnant.
\end{abstract}

\begin{keywords}
supernovae -- shocks -- supernova remnants -- hydrodynamics
\end{keywords}

%%%%%%%%%%%%%%%%%%%%%%%%%%%%%%%%%%%%%%%%%%%%%%%%%%

%%%%%%%%%%%%%%%%% BODY OF PAPER %%%%%%%%%%%%%%%%%%

\section{Introduction}

The supernova remnant (SNR) Pa 30 \citep{2014apn6.confE..48K, 2020A&A...644L...8O, 2021ApJ...918L..33R, 2023ApJ...944..120L} presents a challenging puzzle.  It was first discovered to be a supernova remnant in 2020, with the first X-ray observations of the object \citep{2020A&A...644L...8O}.  Recent imaging of Sulfur emission lines have discovered that its morphology is strikingly unique, with long, narrow, spikey filaments pointing almost perfectly radially outward, and seemingly distributed uniformly in a sphere \citep{2023ApJ...945L...4F}.  Generating such a structure naturally is nontrivial; Rayleigh-Taylor instability (RTI) can generate fingers pointing outward, but typically this introduces a shear flow which generates turbulence via Kelvin-Helmholtz instability (KHI), and the fingers immediately turn over and form bent mushroom-like structures, which no longer point radially outward \citep{1992ApJ...392..118C}.  Thus, RTI will typically generate the more clumpy, cauliflower-like structures seen in SNRs such as Tycho \citep{2006ApJ...645.1373B, 2013MNRAS.429.3099W}.

A strong, fast wind has been inferred from observations of the central object at the center of Pa 30, and it has been suggested that this wind might be responsible for blowing out this radial structure \citep{2023ApJ...945L...4F}.  This interpretation faces two major issues: first, wind-driven Rayleigh-Taylor instabilities suffer from this same KHI cascade into turbulence.  Secondly, the morphology RTI generates in the wind-driven case is very different from that seen in Pa 30.  The most prominent example of wind-driven anisotropies is the Crab nebula \citep{2017ApJ...840...82D}.  The morphology of the Crab is generated by the pulsar wind; in the crab, RTI drives fingers that point inward rather than outward, blowing large bubble-like structures in the remnant, leaving an intricate web of transverse filaments.  Similar morphologies to the Crab are occasionally seen in planetary nebulae such as NGC 3132 \citep{2022ApJ...936L..14P}, which are also blown out by winds.

Invoking a ``clumpy" ejecta or circumstellar medium (CSM) also doesn't solve this problem.  First, it has been shown in many studies that clumps in the CSM do not affect the structure of the remnant \citep{1992ApJ...392..118C, 2013MNRAS.429.3099W, 2022ApJ...940L..28P, 2023ApJ...956..130M}; the forward shock is a stabilizing structure that flattens out external clumps efficiently.  Starting with extremely small and dense clumps in the ejecta might work, similar to what gives rise to very fast radial structures in remnants such as Cas A \citep{2024arXiv240102477M}.  However, these clumps would have to be distributed almost perfectly uniformly in a sphere throughout the ejecta, and all at the same radial velocity; it is not clear what mechanism would generate such an ideal arrangement of dense clumps.

It is also possible that these radial filaments are somehow generated from the ionizing radiation of the compact object at the center of the remnant.  This seems plausible, though to our knowledge a specific mechanism for generating the anisotropy via stellar radiation has not yet been proposed.

In this letter, we propose a mechanism that naturally would give rise to long, narrow radial filaments in a supernova remnant, and we perform 2D numerical hydrodynamics calculations to demonstrate as a proof of concept that it can give qualitative agreement with the observed morphology of Pa 30.

The idea is to drive RTI in the ejecta-CSM interaction, but to suppress KHI via efficient line cooling in the ejecta.  This results in very long, narrow filaments pointing almost perfectly radially outward, uniformly distributed over the sphere.  We do not speculate here what atomic transition might be responsible for this efficient cooling, but we predict that if this is the mechanism responsible, broadband spectra of the ejecta in Pa 30 would reveal very strong emission lines; strong enough to cool the ejecta significantly and modify the dynamics.

We additionally make further ``smoking gun" predictions regarding the shape of the forward shock and the velocity structure in the filaments, which should be measurable in follow-up observations, thus making this idea falsifiable in the near future.  Another advantage to these additional predictions is they would help to distinguish between mechanisms of a hydrodynamical origin from mechanisms involving stellar radiation from the central source.

This letter is structured as follows:  in section \ref{sec:num} we briefly summarize our numerical setup and cooling term.  In section \ref{sec:results} we present the resulting remnant morphology with and without this cooling term, measuring the temperature and velocity information in the filaments.  In section \ref{sec:sum}, we summarize and discuss our predictions for future observations of Pa 30.

\section{Numerical Method}
\label{sec:num}

Hydrodynamics calculations are carried out using the \texttt{JET} code \citep{2011ApJS..197...15D, 2013ApJ...775...87D}.  \texttt{JET} uses a dynamically shearing radial mesh that follows the outflowing gas, efficiently maintaining contact discontinuities and allowing for expansion over a large range of scales.

Initial conditions are given by an exponential density structure for the ejecta and a constant-density circumstellar medium (CSM):

\begin{equation}
    \rho_{\rm ej} = \frac{M_0}{8 \pi v_0^3 t^3} e^{-r/v_0 t}
\end{equation}

\begin{equation}
    \rho_{\rm CSM} = \rm const,
\end{equation}
with $v_0 = \sqrt{E/6 M_0}$.  Velocity in the ejecta is taken to be homologous in the ejecta and zero in the CSM:

\begin{equation}
    \vec v_{\rm ej} = \vec r/t
\end{equation}

\begin{equation}
    \vec v_{\rm CSM} = 0
\end{equation}

The gas is chosen to be very cold everywhere initially.  The calculation is run over a long dynamic range in time, from $t = 10^{-6} T_{\rm sweep}$ to $t = T_{\rm sweep}$, with 

\begin{equation}
    T_{\rm sweep} \equiv \sqrt{\frac{M_0}{E}} \left( \frac{M_0}{\rho_{\rm CSM}} \right)^{1/3}.
\end{equation}

This is roughly the time it takes the ejecta to sweep up its own mass in the CSM.  Additionally a passive scalar, $X$ is evolved with the flow, to discriminate ejecta from CSM.  $X = 1$ in the ejecta and $X = 0$ in the CSM.  Finally, a cooling sink term is added to the energy equation, with the following form:

\begin{equation}
    S_E = - n X \frac{ \Gamma h \nu }{ 1 + e^{h \nu/k T}} = - \rho X \frac{K}{1 + {\rm exp}(C \rho/P)}.
\end{equation}
where $\hbar \omega$ is the transition energy, $n$ is the number density, and $\Gamma$ is the decay rate of the excited state.  As this study will be agnostic to the specifics of the emission mechanism, we varied the values of $K$ and $C$ until cooling was sufficient to produce the structures observed.  In a follow-up study we will study this parameter space more systematically.

In practice, we found $K = 1000$, $C = 0.01$ (in code units where $E = M_0 = \rho_{\rm CSM} = 1$) gave sufficient cooling to impact the RTI.  In section \ref{sec:sum} we will discuss what these values imply in physical units.

\section{Results}
\label{sec:results}

\begin{figure}
\includegraphics[width=3.3in]{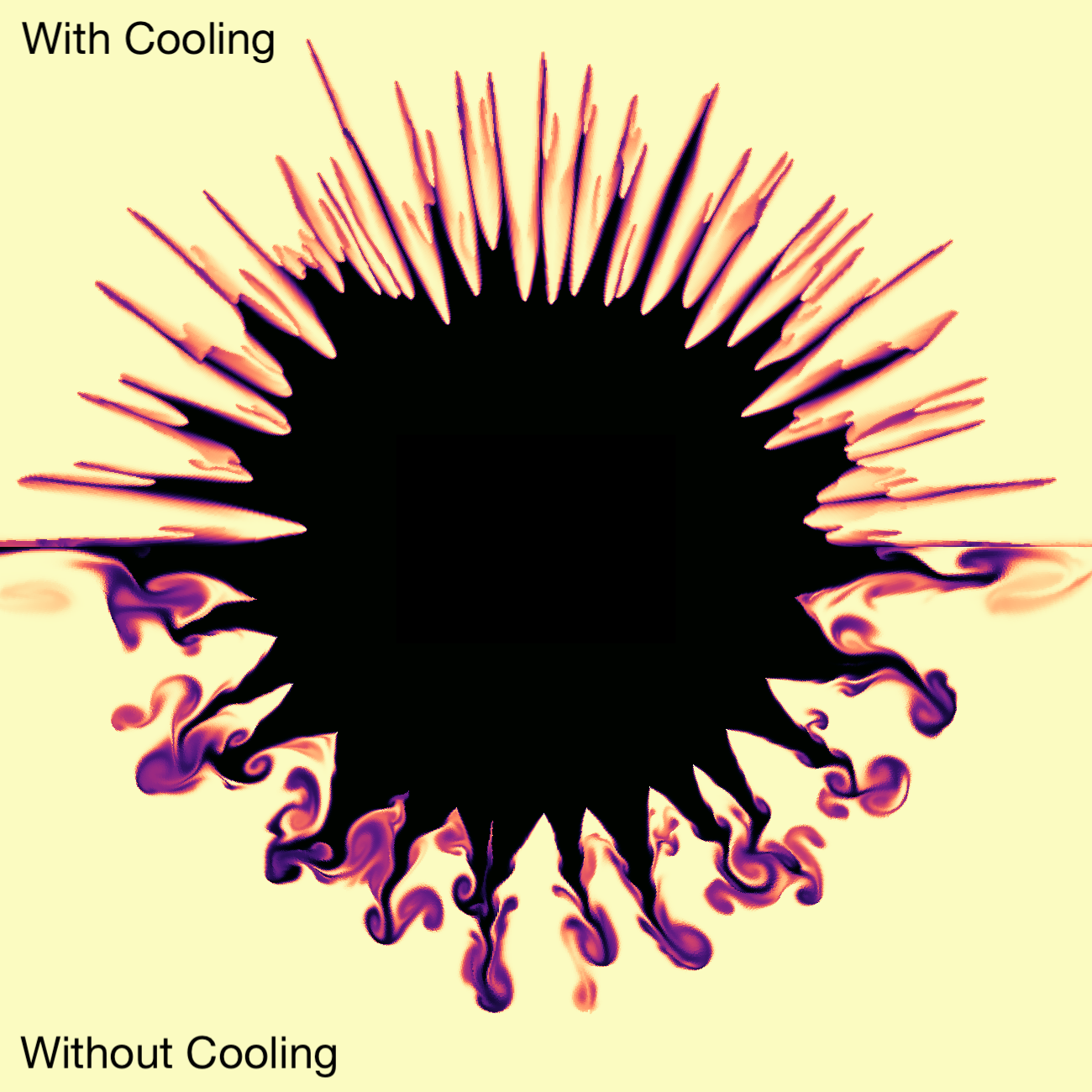}
\caption{Passive scalar X denoting composition (ejecta vs. CSM) at $t = T_{\rm sweep}$.  Upper half employs cooling and lower half included no cooling.}
\label{fig:pretty}
\end{figure}

Figure \ref{fig:pretty} shows the passive scalar $X$ in our calculation at $t = T_{\rm sweep}$, when large filamentary structures become most prominent.  Two calculations were run, one with cooling turned on and the other with cooling turned off, to demonstrate the impact this has on the structure of the remnant.

RTI is present in both cases, but it is apparent that the RTI fingers exhibit very different morphology when the cooling is highly efficient.  Because the ejecta is cooled but not the CSM, pressure support is reduced within the RTI fingers, which are then compressed by the CSM, suppressing KHI and allowing RTI fingers to grow radially without turning over.  Thus, the instability does not form the typical characteristic mushroom-like structures and cascade into turbulence.  Instead, the fingers remain long, narrow and pointed radially.

If the filaments were not perfectly radial, then the projection from 3D to the observed 2D image would result in a much less ordered remnant than is observed.  It is not good enough to provide a mechanism that generates long filaments, these filaments must not bend or turn over.  Thus, in our calculations, it is an important result that the fingers are pointed precisely in the radial direction when cooling is included.

\section{Conclusions}
\label{sec:sum}

\subsection{Interpretation of Pa 30's Nebular Structure}

In order to generate this highly ordered structure hydrodynamically via this mechanism, one needs very strong cooling in the shocked ejecta.  Our model assumed emission line cooling, but it is possible that some other cooling mechanism could be responsible.  This necessitated a cooling coefficient of $K = 1000$ in our code units.  In physical units, this corresponds to a total luminosity of

\begin{equation}
    L \sim 10^{41} ~{\rm erg/s} \left( \frac{M_X}{0.1 M_\odot} \right) \left( \frac{v_{\rm ej}}{1000 ~{\rm km/s}} \right)^2 \left( \frac{t}{850 ~\rm y} \right)^{-1}
\end{equation}

where $M_X$ is the total mass of the emitting material.

However, in practice, the cooling rate does not attain this value; in our models, cooling removed about 70\% of the ejecta energy, which reduced the temperature enough to affect the value of the cooling term.  It would be more appropriate to say the above formula implies a very short cooling timescale.  To be more precise about the amount of cooling in our calculations, we summed up our cooling sink term across the domain at our final checkpoint, and found

\begin{equation}
    L = 1.5 \times 10^{38} ~{\rm erg/s} \left( \frac{M}{M_\odot} \right) \left( \frac{v_{\rm ej}}{1000 ~{\rm km/s}} \right)^2 \left( \frac{t}{850 ~\rm y} \right)^{-1}.
    \label{eqn:lum}
\end{equation}

It should be noted that we did not investigate parameter space in detail, and in particular we chose to look at morphology at a dynamical age of $t=T_{\rm sweep}$, the time when the swept-up mass is comparable to the ejecta mass.  The true dynamical age of the remnant might be much younger. 

We propose that some strong emission line (or lines) could exist in the ejecta, with enough luminosity to significantly cool the gas in an expansion time.  Equation (\ref{eqn:lum}) gives a ballpark estimate for the necessary luminosity, but we have not explored parameter space thoroughly, so it might be possible to generate these structures with significantly less cooling.

Strong Sulfur lines have already been seen in the ejecta \citep{2023ApJ...945L...4F}, but it is possible that this Sulfur is mixed with other intermediate-mass elements that cool even more efficiently.  The composition in the shocked region will depend sensitively on the temperature achieved during the explosion, and the completeness (or incompleteness) of the burning process during the supernova.  \cite{2023arXiv230414669K} argue that very strong cooling should be exhibited in the shocked ejecta within the wind termination shock.  \cite{2020A&A...644L...8O} also observed very strong Magnesium and Neon lines in the shocked ejecta, suggesting that this remnant has an unusual composition relative to typical Ia remnants.

\subsection{Why Pa 30 and not other remnants?}

The morphology of Pa 30 is unique; if this unusual structure can be so naturally generated, why don't we see it in all remnants?

We consider two possible reasons: first, Pa 30 has much slower ejecta velocities than other comparable thermonuclear remnants such as Tycho.  This could mean that it is easier for cooling to make an impact, i.e. for cooling to remove a significant fraction of the ejecta kinetic energy.  We have found that reducing the strength of the cooling term relative to the energy of the supernova causes the morphology to look closer to the ``without cooling'' case.  Equation (\ref{eqn:lum}) gives an estimate for the cooling luminosity necessary to affect the morphology; this necessary luminosity scales as $v_{\rm ej}^2$.  It is possible that Pa 30 is a special case of a supernova with low kinetic energy relative to the amount of energy radiated in an expansion time.

Secondly, it could be that the composition of Pa 30 is unique.  It was a thermonuclear supernova with slower expansion velocities than Tycho, which suggests it released significantly less nuclear binding energy per unit mass.  If the supernova associated with Pa 30 exhibited incomplete burning, then there could be a larger quantity of intermediate mass elements than is seen in the ejecta of typical Ia's.  Additionally, one only needs these strong emission lines in the {\em shocked} ejecta; it could be that Pa 30 has comparable yields to other supernovae, but an unusual distribution of those elements, with some significant quantity of an efficiently-cooling intermediate mass element in the shocked region.

The X-ray spectra found by \cite{2020A&A...644L...8O} suggest that the composition is indeed unlike typical Ia remnants, with very strong lines of Magnesium and Neon, possibly suggesting that the initial explosion could have been triggered by an unusual progenitor, such as an Oxygen-Neon white dwarf.

\subsection{Other possible mechanisms for Anisotropy}

We have discussed other hydrodynamical mechanisms in the introduction, but we will flesh out a few of the key points here.

First, we address the question of winds.  Observations show that the central object of Pa 30 exhibits very fast winds ($\sim 16,000$ km/s! \cite{2023ApJ...945L...4F}).  It is reasonable to ask whether these filaments are somehow imparted or impacted by this wind.

However, when a wind is blown into a supernova's ejecta, the morphology one attains is not that of long filamentary radial structures.  It is actually the opposite; large, transverse filaments forming an intricate web surrounding a hot bubble being blown out by the wind (the Crab nebula is a canonical example).  Moreover, such a bubble does not advance with the same speed as the wind velocity; one can estimate the termination shock has an expansion velocity of

\begin{equation}
    v_{\rm term}(t) \sim ( L_w t / M_{\rm ej} )^{1/5} v_{\rm ej}^{3/5},
\end{equation}

where $L_w$ is the luminosity of the wind.  Assuming $L_w = \frac12 \dot M v_w^2$, this gives

\begin{equation}
    v_{\rm term}(t) \sim ( \dot M t / M_{\rm ej} )^{1/5} v_{\rm ej}^{3/5} v_w^{2/5}.
\end{equation}

or in terms of $v_{\rm term}/v_{\rm ej}$,

\begin{equation}
    v_{\rm term}/v_{\rm ej} \sim ( \dot M t / M_{\rm ej} )^{1/5} (v_w/v_{\rm ej})^{2/5}.
\end{equation}

As $v_w / v_{\rm ej} \sim 10$ \citep{2021ApJ...918L..33R, 2023ApJ...945L...4F}, the termination shock will not overtake the ejecta until $\dot M t \sim 0.01 M_{\rm ej}$, which would necessitate  $\dot M \sim 10^{-5} M_{\sun}$/yr, which would be a very high mass loss rate for the compact central object.  \cite{2019Natur.569..684G} and \cite{2023ApJ...944..120L} report a mass loss rate that is an order of magnitude lower, $\sim 10^{-6} M_{\sun}$/yr.  Thus, unless the mass loss rate in the wind is much more extreme than inferred from observations, the bubble inflated by the wind has not yet reached the size of the Pa 30 nebula.  In fact, Pa 30 features a bright X-ray region at much smaller velocities (with a maximum size of 0.018 pc) which would be consistent with a wind-powered shock that has not yet expanded out to the filaments \citep{2023arXiv230414669K}.

Secondly, we address the possibility of ``clumpy" ejecta or CSM, which have been tested in many studies; clumps in the CSM never impart strong anisotropies onto the remnant \citep{1992ApJ...392..118C, 2013MNRAS.429.3099W, 2022ApJ...940L..28P, 2023ApJ...956..130M}.  Certainly it does not impart features as distinctive as those long, filamentary radial structures in Pa 30.

What about clumps in the ejecta?  Ejecta clumps could generate these features, but it begs the question of where those clumps came from in the first place, and why they are ordered so symmetrically around the supernova, all at the same velocity.  Actually, one could make the semantic argument that our mechanism generates clumps in the ejecta, which then produce these radial filamentary structures.  So, clumps in the ejecta could indeed generate Pa 30's morphology, but it just raises further questions as to where these clumps come from, which we have arguably provided a solution for in this study.

The filaments point radially outward, in the direction of the supernova expansion velocity.  This is also the direction away from the central object, which is extremely hot with a high radiative luminosity.  It is possible this intense radiation is capable of sculpting the structure of the remnant in some way.  We do not speculate on any particular mechanism, except to say that a radiative process like this would predict very different velocity structure and shock structure than a hydrodynamical process, so determining whether the mechanism is radiative or hydrodynamical in nature should be possible with follow-up observations.

\subsection{Prediction \#1: Strong Emission Lines in the Filaments}

If there is strong cooling in Pa 30, it has to come out in radiation at some observable wavelength (unless the cooling is caused by cosmic ray acceleration, but we do not consider that here).  In particular, if the cooling is specific to the ejecta, necessitating a specific emission mechanism depending on the composition, one should expect to see very strong emission lines in some band.  The filaments of Pa 30 trace strong observed Sulfur emission lines, but the cooling process might not occur through sulfur; there may be some other intermediate-mass element mixed with the sulfur (e.g. Silicon, Magnesium, or Calcium) that could cool the remnant even more efficiently.  Whatever the line, it should be responsible for removing a significant amount of thermal energy in the remnant, so it is likely to stand out in a broadband spectrum taken of the filaments.

\subsection{Prediction \#2: Highly Corrugated Forward Shock}

It is known that cooling can lead to a corrugated forward shock.  For example, cosmic ray cooling in typical SNRs (like Tycho) allows RTI fingers to catch up with the forward shock and distort it.  The cooling in this case is much more extreme than the cosmic ray cooling in Tycho; in our model, the forward shock wraps tightly around the RTI fingers, essentially forming a tight ``wake" that the fingers generate as they collide with the CSM.  As a result, the forward shock structure is predicted to be highly anisotropic on small scales (i.e. on the size of the filaments).  This would not affect low-resolution X-ray images of Pa 30, but should be observable with sufficiently long-exposure Chandra imaging.

\subsection{Prediction \#3: Nearly-Ballistic Filament Velocities}

In our model, the filaments are efficiently cooled into narrow high-density fingers.  Because these fingers are so dense and the shocks are relatively cold, the fingers exhibit very little deceleration when colliding with the surrounding medium.  Thus, the velocity structure is nearly ballistic and homologous; we measured the velocity along these fingers and found

\begin{equation}
    v \approx 0.93 ~r/t.
\end{equation}

Thus, the velocity of the filaments are found to be about 93\% of the ballistic expansion velocity (i.e. very nearly ballistic, and not significantly decelerated by interaction with the CSM).  Note that this measurement could strongly distinguish this mechanism from the formation of structure in a wind.  If the filaments are part of a wind, one would expect either a constant large velocity along the fingers if the filaments had not encountered any ejecta, or a velocity that was greater than that predicted by homologous expansion ($v>r/t$), if the wind had accelerated the ejecta.

Detailed follow-up observations of the filament velocity as a function of radius should be able to test whether the filaments are ballistic or if they have been significantly decelerated by shocks or accelerated by a wind.  If they are nearly ballistic (i.e. the velocity along the filament is less than but very close to $r/t$), that is strong evidence that significant cooling has occurred.  \cite{2023ApJ...945L...4F} has reported variable velocity dispersion along the length of a filament, consistent with a velocity proportional to distance from the center; a more precise measurement of the velocity as a function of radius should be possible in the near future.

\section*{Data Availability Statement}

The data underlying this article will be shared on reasonable request to the corresponding author.  All hydrodynamic caluclations were carried out by the \texttt{JET} code, which is publicly available at the following URL: https://github.com/chomps/Jet

\section*{Acknowledgements}

PD is supported by the National Science Foundation under grant No. AAG-2206299.  Numerical calculations were performed using the \texttt{Petunia} cluster at Purdue University.

%\bibliography{refs}

\begin{thebibliography}{}
\makeatletter
\relax
\def\mn@urlcharsother{\let\do\@makeother \do\$\do\&\do\#\do\^\do\_\do\%\do\~}
\def\mn@doi{\begingroup\mn@urlcharsother \@ifnextchar [ {\mn@doi@} {\mn@doi@[]}}
\def\mn@doi@[#1]#2{\def\@tempa{#1}\ifx\@tempa\@empty \href {http://dx.doi.org/#2} {doi:#2}\else \href {http://dx.doi.org/#2} {#1}\fi \endgroup}
\def\mn@eprint#1#2{\mn@eprint@#1:#2::\@nil}
\def\mn@eprint@arXiv#1{\href {http://arxiv.org/abs/#1} {{\tt arXiv:#1}}}
\def\mn@eprint@dblp#1{\href {http://dblp.uni-trier.de/rec/bibtex/#1.xml} {dblp:#1}}
\def\mn@eprint@#1:#2:#3:#4\@nil{\def\@tempa {#1}\def\@tempb {#2}\def\@tempc {#3}\ifx \@tempc \@empty \let \@tempc \@tempb \let \@tempb \@tempa \fi \ifx \@tempb \@empty \def\@tempb {arXiv}\fi \@ifundefined {mn@eprint@\@tempb}{\@tempb:\@tempc}{\expandafter \expandafter \csname mn@eprint@\@tempb\endcsname \expandafter{\@tempc}}}

\bibitem[\protect\citeauthoryear{{Badenes}, {Borkowski}, {Hughes}, {Hwang}  \& {Bravo}}{{Badenes} et~al.}{2006}]{2006ApJ...645.1373B}
{Badenes} C.,  {Borkowski} K.~J.,  {Hughes} J.~P.,  {Hwang} U.,   {Bravo} E.,  2006, \mn@doi [\apj] {10.1086/504399}, \href {https://ui.adsabs.harvard.edu/abs/2006ApJ...645.1373B} {645, 1373}

\bibitem[\protect\citeauthoryear{{Chevalier}, {Blondin}  \& {Emmering}}{{Chevalier} et~al.}{1992}]{1992ApJ...392..118C}
{Chevalier} R.~A.,  {Blondin} J.~M.,   {Emmering} R.~T.,  1992, \mn@doi [\apj] {10.1086/171411}, \href {https://ui.adsabs.harvard.edu/abs/1992ApJ...392..118C} {392, 118}

\bibitem[\protect\citeauthoryear{{Dubner}, {Castelletti}, {Kargaltsev}, {Pavlov}, {Bietenholz}  \& {Talavera}}{{Dubner} et~al.}{2017}]{2017ApJ...840...82D}
{Dubner} G.,  {Castelletti} G.,  {Kargaltsev} O.,  {Pavlov} G.~G.,  {Bietenholz} M.,   {Talavera} A.,  2017, \mn@doi [\apj] {10.3847/1538-4357/aa6983}, \href {https://ui.adsabs.harvard.edu/abs/2017ApJ...840...82D} {840, 82}

\bibitem[\protect\citeauthoryear{{Duffell} \& {MacFadyen}}{{Duffell} \& {MacFadyen}}{2011}]{2011ApJS..197...15D}
{Duffell} P.~C.,  {MacFadyen} A.~I.,  2011, \mn@doi [\apjs] {10.1088/0067-0049/197/2/15}, \href {https://ui.adsabs.harvard.edu/abs/2011ApJS..197...15D} {197, 15}

\bibitem[\protect\citeauthoryear{{Duffell} \& {MacFadyen}}{{Duffell} \& {MacFadyen}}{2013}]{2013ApJ...775...87D}
{Duffell} P.~C.,  {MacFadyen} A.~I.,  2013, \mn@doi [\apj] {10.1088/0004-637X/775/2/87}, \href {https://ui.adsabs.harvard.edu/abs/2013ApJ...775...87D} {775, 87}

\bibitem[\protect\citeauthoryear{{Fesen}, {Schaefer}  \& {Patchick}}{{Fesen} et~al.}{2023}]{2023ApJ...945L...4F}
{Fesen} R.~A.,  {Schaefer} B.~E.,   {Patchick} D.,  2023, \mn@doi [\apjl] {10.3847/2041-8213/acbb67}, \href {https://ui.adsabs.harvard.edu/abs/2023ApJ...945L...4F} {945, L4}

\bibitem[\protect\citeauthoryear{{Gvaramadze}, {Gr{\"a}fener}, {Langer}, {Maryeva}, {Kniazev}, {Moskvitin}  \& {Spiridonova}}{{Gvaramadze} et~al.}{2019}]{2019Natur.569..684G}
{Gvaramadze} V.~V.,  {Gr{\"a}fener} G.,  {Langer} N.,  {Maryeva} O.~V.,  {Kniazev} A.~Y.,  {Moskvitin} A.~S.,   {Spiridonova} O.~I.,  2019, \mn@doi [\nat] {10.1038/s41586-019-1216-1}, \href {https://ui.adsabs.harvard.edu/abs/2019Natur.569..684G} {569, 684}

\bibitem[\protect\citeauthoryear{{Ko} et~al.,}{{Ko} et~al.}{2023}]{2023arXiv230414669K}
{Ko} T.,  et~al., 2023, \mn@doi [arXiv e-prints] {10.48550/arXiv.2304.14669}, \href {https://ui.adsabs.harvard.edu/abs/2023arXiv230414669K} {p. arXiv:2304.14669}

\bibitem[\protect\citeauthoryear{{Kronberger} et~al.,}{{Kronberger} et~al.}{2014}]{2014apn6.confE..48K}
{Kronberger} M.,  et~al., 2014, in {Morisset} C.,  {Delgado-Inglada} G.,   {Torres-Peimbert} S.,  eds, Asymmetrical Planetary Nebulae VI Conference. p.~48

\bibitem[\protect\citeauthoryear{{Lykou}, {Parker}, {Ritter}, {Zijlstra}, {Hillier}, {Guerrero}  \& {Le D{\^u}}}{{Lykou} et~al.}{2023}]{2023ApJ...944..120L}
{Lykou} F.,  {Parker} Q.~A.,  {Ritter} A.,  {Zijlstra} A.~A.,  {Hillier} D.~J.,  {Guerrero} M.~A.,   {Le D{\^u}} P.,  2023, \mn@doi [\apj] {10.3847/1538-4357/acb138}, \href {https://ui.adsabs.harvard.edu/abs/2023ApJ...944..120L} {944, 120}

\bibitem[\protect\citeauthoryear{{Mandal}, {Duffell}, {Polin}  \& {Milisavljevic}}{{Mandal} et~al.}{2023}]{2023ApJ...956..130M}
{Mandal} S.,  {Duffell} P.~C.,  {Polin} A.,   {Milisavljevic} D.,  2023, \mn@doi [\apj] {10.3847/1538-4357/acf9fb}, \href {https://ui.adsabs.harvard.edu/abs/2023ApJ...956..130M} {956, 130}

\bibitem[\protect\citeauthoryear{{Milisavljevic} et~al.,}{{Milisavljevic} et~al.}{2024}]{2024arXiv240102477M}
{Milisavljevic} D.,  et~al., 2024, \mn@doi [arXiv e-prints] {10.48550/arXiv.2401.02477}, \href {https://ui.adsabs.harvard.edu/abs/2024arXiv240102477M} {p. arXiv:2401.02477}

\bibitem[\protect\citeauthoryear{{Oskinova}, {Gvaramadze}, {Gr{\"a}fener}, {Langer}  \& {Todt}}{{Oskinova} et~al.}{2020}]{2020A&A...644L...8O}
{Oskinova} L.~M.,  {Gvaramadze} V.~V.,  {Gr{\"a}fener} G.,  {Langer} N.,   {Todt} H.,  2020, \mn@doi [\aap] {10.1051/0004-6361/202039232}, \href {https://ui.adsabs.harvard.edu/abs/2020A&A...644L...8O} {644, L8}

\bibitem[\protect\citeauthoryear{{Polin}, {Duffell}  \& {Milisavljevic}}{{Polin} et~al.}{2022}]{2022ApJ...940L..28P}
{Polin} A.,  {Duffell} P.,   {Milisavljevic} D.,  2022, \mn@doi [\apjl] {10.3847/2041-8213/aca28b}, \href {https://ui.adsabs.harvard.edu/abs/2022ApJ...940L..28P} {940, L28}

\bibitem[\protect\citeauthoryear{{Pontoppidan} et~al.,}{{Pontoppidan} et~al.}{2022}]{2022ApJ...936L..14P}
{Pontoppidan} K.~M.,  et~al., 2022, \mn@doi [\apjl] {10.3847/2041-8213/ac8a4e}, \href {https://ui.adsabs.harvard.edu/abs/2022ApJ...936L..14P} {936, L14}

\bibitem[\protect\citeauthoryear{{Ritter}, {Parker}, {Lykou}, {Zijlstra}, {Guerrero}  \& {Le D{\^u}}}{{Ritter} et~al.}{2021}]{2021ApJ...918L..33R}
{Ritter} A.,  {Parker} Q.~A.,  {Lykou} F.,  {Zijlstra} A.~A.,  {Guerrero} M.~A.,   {Le D{\^u}} P.,  2021, \mn@doi [\apjl] {10.3847/2041-8213/ac2253}, \href {https://ui.adsabs.harvard.edu/abs/2021ApJ...918L..33R} {918, L33}

\bibitem[\protect\citeauthoryear{{Warren} \& {Blondin}}{{Warren} \& {Blondin}}{2013}]{2013MNRAS.429.3099W}
{Warren} D.~C.,  {Blondin} J.~M.,  2013, \mn@doi [\mnras] {10.1093/mnras/sts566}, \href {https://ui.adsabs.harvard.edu/abs/2013MNRAS.429.3099W} {429, 3099}

\makeatother
\end{thebibliography}

%%%%%%%%%%%%%%%%%%%%%%%%%%%%%%%%%%%%%%%%%%%%%%%%%%
%\section*{Data Availability}

%The inclusion of a Data Availability Statement is a requirement for articles published in MNRAS. Data Availability Statements provide a standardised format for readers to understand the availability of data underlying the research results described in the article. The statement may refer to original data generated in the course of the study or to third-party data analysed in the article. The statement should describe and provide means of access, where possible, by linking to the data or providing the required accession numbers for the relevant databases or DOIs.


\bibliographystyle{mnras}

% Don't change these lines
\bsp	% typesetting comment
\label{lastpage}
\end{document}